\documentclass[runningheads,citeauthoryear]{apinv}
\usepackage{epsfig,cite,graphics}
\usepackage[T2A]{fontenc}
\usepackage[cp1251]{inputenc}

\begin{document}

\title{SOLUTION OF NEWLY OBSERVED TRANSIT
OF THE EXOPLANET HAT-P-24B:
NO TTV AND TDV SIGNALS\,\thanks{based on NAO Rozhen observations}}
\titlerunning{SOLUTION OF NEWLY OBSERVED TRANSIT  \\
OF HAT-P-24B\\
 }
\author{Diana Kjurkchieva\inst{1}, Dinko Dimitrov\inst{2} and Sunay Ibryamov\inst{1,2}}
\authorrunning{D. Kjurkchieva, D. Dimitrov, S. Ibryamov}
\tocauthor{D. Kjurkchieva}
% Command tocautor{} is used by the Latex to give author names
% to the Contents of the volume (automatically generated)
\institute{Department of Physics, University of Shumen, 115 Universitetska Str., Shumen, Bulgaria
    \and Institute of Astronomy and NAO, Bulgarian Academy of Sciences, 72 Tsarigradsko Shose, Sofia, Bulgaria\newline
     \email{d.kyurkchieva@shu-bg.net};
     \email{dinkopd@gmail.com};
 \email{sibryamov@shu.bg}  }
\papertype{Submitted on xx.xx.xxxx; Accepted on xx.xx.xxxx}
% Papertype can be "Research report", "Review", "Invited lecture", "Conference talk",
% "Conference poster", "Lecture at scientific seminar", "Summary of dissertation",  etc.
\maketitle

\begin{abstract}
We present photometric observations of transit of the exoplanet
HAT-P-24b using the Rozhen 2 m telescope. Its solution gives
relative stellar radius $r_s$=0.1304 ($a/R_s$ = 7.669), relative
planet radius $r_p$=0.01304 and orbital inclination of 90$^\circ$.
The calculated planet radius is $R_p$ = 1.316 R$_J$ and
corresponds to planet density of $\rho_p$ = 0.37 g cm$^{-3}$. Our
parameter values are between those of the previous two solutions.
We did not find evidences of TTV and TDV signals of HAT-P-24b.
\end{abstract}


\section*{Introduction}

Photometric observations of transiting exoplanets (TEPs) provide
accurate relative sizes of their configurations, which are
important for understanding their formation and evolution
(Charbonneau et al. 2000). The monitoring of transiting exoplanet
allows to improve its physical parameters as well as to study
existence of other planets in the system by analysis of the
transit timing variation (TTV) and transit duration variation
(TDV) (Agol et al. 2005; Holman $\&$ Murray 2005; Kipping
2009a,b). Moreover, the transits provide a possibility to learn
the oblateness of the planets (Seager $\&$ Hui 2002; Carter $\&$
Winn 2010) and to make thermal mapping of their surfaces (Knutson
et al. 2007).

During the last two decades the number of discovered and studied
exoplanets rapidly increased as a consequence mostly of automated
surveys as SuperWASP (Street et al. 2004), CoRoT (Baglin et al.
2007), HATNet (Bakos et al. 2004), TrES (Alonso et al. 2004), XO
(McCullough et al. 2005), KELT (Pepper et al. 2007), OGLE (Udalski
et al. 1998) and \emph{Kepler} (Koch et al. 2010). The
considerable part of the newly discovered exoplanets are hot
Jupiters. These are a class of extrasolar planets whose
characteristics are similar to Jupiter, but which have higher
surface temperatures because they orbit very close (0.015--0.1 AU)
to their parent (F, G and K-type) stars. That is why the
probability to observe transits of hot Jupiters is bigger compared
to other known types of planets. Morerover, they are the easiest
extrasolar planets for confirmation via the radial-velocity
method, because the induced oscillations in the star motion are
relatively large and rapid. Hot Jupiters are gas giants with low
density although there are examples which are denser than Jupiter:
WASP-18 (7 times the density of Jupiter, Southworth et al. 2009)
and HAT-P-2 (9 times the density of Jupiter, Pal et al. 2010).

%The temperature difference between their two sides is relatively low that implies high-speed winds distributing the heat from the day side to the night side of the planet surface.

It is supposed that hot Jupiters have migrated to their present
positions after their formation at a distance from the star beyond
the frost line. The most investigations show that they appear to
be alone or accompanied by other planetary bodies on wide orbits
(Steffen et al. 2012) but the studies of the \emph{Kepler}
candidates for hot planets reveal that some of them exhibit
periodic variations in transit timing (Szabo et al. 2013). These
multiple-planet systems may provide some constraints on planet
formation and migration theories (Podlewska $\&$ Szuszkiewicz
2009). Moreover, the distribution of planet period ratios shows
strange clumping around the low-order resonances (Fabrycky et al.
2014). On the other hand the loneliness of the gas giants is in
favor of inward migration theories of massive outer planets
through planet-planet scattering caused by mutual dynamical
perturbations (Weidenschilling $\&$ Marzari 1996; Rasio $\&$ Ford
1996).  That is why the study of hot Jupiters is important area of
the modern astrophysics.

HAT-P-24b was discovered as TEP on the base of HATNet
(Hungarian-made Automated Telescope Network) photometric
observations and confirmed by Keck spectral observations (Kipping
et al. 2010). It orbits the F8V star GSC 0774--01441 (mass of 1.191
M$\odot$, radius of 1.317 R$\odot$, effective temperature 6373 K,
metallicity of [Fe/H] = --0.16) with a period P = 3.3552464 days
on slightly eccentric orbit ($e$ = 0.052) with semiaxis \emph{a} =
0.04641 AU (Kipping et al. 2010). HAT-P-24b is classified as an
inflated hot Jupiter with a mass of 0.681 M$_J$ and radius of
1.243 R$_J$ (mean density $\rho_p$ = 0.439 g cm$^{-3}$).

Wang et al. (2013) obtained three complete transit light curves of
HAT-P-24b in 2010--2012.They derived slightly bigger orbital period
and larger planet radius $R_p$ = 1.364 R$_J$.

We carried out observations and study of HAT-P-24b in order to
obtain new transit solution and to search for TTV and/or TDV
signal as indications of presence of other bodies in the system.

\section*{OBSERVATIONS AND DATA REDUCTION}

Our CCD photometric observations in \emph{R} band were obtained on
Feb 18 2015 at the Rozhen Observatory. We used the 2 m RCC
telescope equipped with focal reducer FoReRo-2 and the CCD camera
VersArray 1300B and (1340 $\times$ 1300
pixels, 20 $\mu$m/pixel, diameter of field of 15 arcmin). The
exposures were 50 s and defocusing was applied. The observations
started around 1 h before the expected beginning of the transit
and ended around 1 h after the event. The average photometric
precision per data point was around 0.0007 mag (the atmospheric
conditions during the second half of the night were not perfect).

\begin{figure}[!htb]
  \begin{center}
    \centering{\epsfig{file=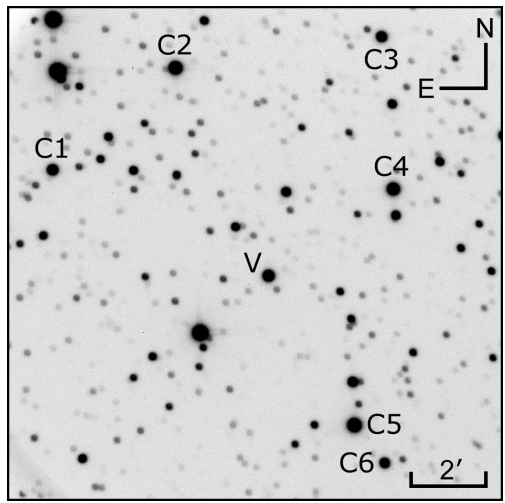, width=0.5\textwidth}}
    \caption[]{The field of HAT-P-24b}
    \label{countryshape}
  \end{center}
\end{figure}
The standard procedures were used for reduction of the photometric
data by \textsc{MaxImDL}. We tested several sets of reduction
parameters and chose the set that gave the most precise photometry
for the stars of similar brightness or brighter than the target.
After carefully selecting of reference stars (Table 1, Fig. 1) we
performed differential aperture photometry by \textsc{MaxImDL}.
The data were cleaned of trends.

The depth of the Rozhen transit (Fig. 2) is around 11.86$\pm$0.28
mmag.

\begin{table}[htb]
  \begin{center}
  \caption{Coordinates and magnitudes of the
target (V) and comparison (C) stars}
\begin{scriptsize}
  \begin{tabular}{cccccc}
\hline\hline
Label  & Star                           & RA (2000)   & DEC (2000)   & R (mag) \\
\hline
V  & HAT-P-24 (2MASS J07151801+1415453) & 07 15 18.01 & +14 15 45.36 & 11.80\\
C1 & 2MASS J07154233+1418402            & 07 15 42.34 & +14 18 40.25 & 12.30\\
C2 & 2MASS J07152847+1421277            & 07 15 28.47 & +14 21 27.74 & 10.90\\
C3 & 2MASS J07150522+1422172            & 07 15 05.22 & +14 22 17.26 & 12.00\\
C4 & 2MASS J07150390+1418079            & 07 15 03.90 & +14 18 17.92 & 11.10\\
C5 & 2MASS J07150839+1411395            & 07 15 08.40 & +14 11 39.53 & 10.70\\
C6 & 2MASS J07150500+1410380            & 07 15 05.01 & +14 10 38.01 & 12.50\\
\hline\hline
  \end{tabular}
  \label{table1}
    \end{scriptsize}
		\end{center}
\end{table}

\begin{figure}
   \centering
   \includegraphics[width=8cm, angle=0]{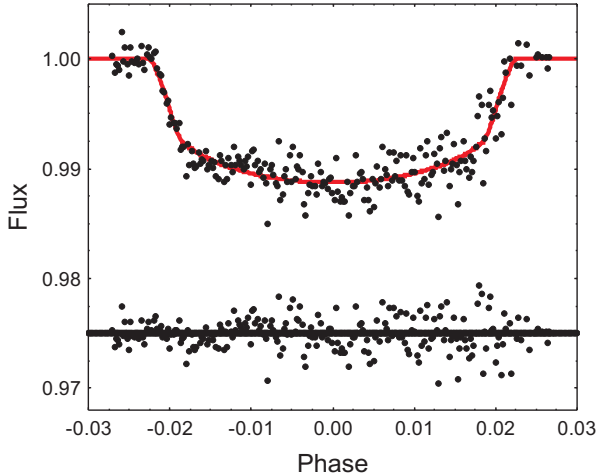}
  % \begin{minipage}[]{85mm}
   \caption{Top: the Rozhen transit of HAT-P-24b and the
synthetic curve corresponding to the best solution; bottom:
the residuals of the fit (shifted vertically by 0.975)}
%\end{minipage}
   \label{Fig3}
   \end{figure}

\section*{MODEL OF THE OBSERVED TRANSIT}

Our observations were modelled using the method of Kjurkchieva et
al. (2013) by the code \textsc{TAC maker 1.1.1} (Kjurkchieva et
al. 2014). It does not use any simplifications of the
configuration (as dark planet, linear planet trajectory on the
stellar disk, etc.) and works with arbitrary
stellar limb-darkening law and planet temperature. Moreover, the
code allows acquisition of the stellar limb-darkening coefficients
from the transit solution and comparison with the theoretical
values (as those of Claret 2004).

We fixed the stellar temperature $T_{s}$ = 6373 K and used the
ephemeris of Kipping et al. (2010). As the first stage of our
solution we adopted linear limb-darkening law with limb-darkening
coefficient corresponding to the stellar temperature according to
the tables of Van Hamme (1993). The adjusted parameters were
relative stellar radius $r_s=R_s/a$ ($a$ is the orbital radius),
relative planet radius $r_p=R_p/a$, orbital inclination \emph{i},
planet temperature $T_{p}$ and the transit time $T_{c}(BJD)$. We
varied them around their values from the solution of Kipping et
al. (2010) to search for minimum of $\chi^2$ (the sum of squares
of the residuals).
%= \sum (f_{obs}-f_{syn})^2$.
%We sorted the obtained solutions by their $\chi^2$ and thus found the best solution.

Finally we tested different limb-darkening laws varying their
coefficients.

The results of our best transit solution corresponding to
quadratic limb-darkening law are given in Table 2. The synthetic
curve is shown in Fig. 2 as continuous line.

% Rs=1.292  Rp=0.1232 Ts=6373 Tp=1000 i=90 , linear law  u1=0.58;
%$R_p$=0.1304    $R_p/R_s$ = 0.1      0.0970$\pm$0.0012
%$R_s$=1.304$\pm$0.001   $a/R_s$ =7.6687                   7.58$\pm$0.35
%Tp=0-600 K
%t0=2455216.97688
%t-center(bjd)=2457072.4281925 +/- 0.000480
%3.3552464   duration 220.32 minute
%a=0.04641 AU

The comparison of our results with the previous solutions of the
target (Table 2) allows us to make several conclusions.

(1) The center of our transit is at $T_{c}(BJD)=2457072.4281925
\pm 0.000480$. Our value (O -- C) = 0.00019 $\pm$ 0.000480 d means
absence of TTV signal. This result supports the same conclusion of Wang
et al. (2013) based on their transits of HAT-P-24b.

(2) Our value of orbital inclination (Table 2) is bigger by around
1.5 $\%$ than that of Wang et al. (2013) but the same as that of
Kipping et al. (2010, Table 6 last column). This value was
necessary to reproduce the depth and shape of the Rozhen transit.

(3) The planet temperature was varied in the reasonable range 0 K
-- 2500 K. We found absolutely the same best fit quality (corresponding to equal minimum value
$\chi^2_{min}$) for planet temperature in the range 0 K -- 1100 K.
It turned out that the transit solution is very insensitive to the
planet temperature: the value of $\chi^2$ for the calculated
planet temperature $T_{eq}$ = 1637 K (from stellar temperature and
planet distance) is only around 1.1$\times 10^{-7}$ $\%$ bigger
than $\chi^2_{min}$ while the value of $\chi^2$
for the planet temperature 2500 K is around 0.015
$\%$ bigger than $\chi^2_{min}$.
%Only the higher sensibility of future observations at longer wavelengths (IR) to the planet emission may provide direct determination of the planet temperature of HAT-P-24b.

(4) The duration of newly-observed transit of 3.672 $\pm$ 0.024 hr
(220.32 $\pm$ 1.47 min) is in the framework of the errors of the
value of Kipping et al. (2010) of 3.653 $\pm$ 0.025 hr (219.18
$\pm$ 1.5 min) and that of Wang et al. (2013) of 3.684 $\pm$ 0.053 hr
(221.04 $\pm$ 3.17 min). Thus, we did not find clear indication of
transit duration variation (TDV). Although the ingress and egress
boundaries of hot Jupiters are harder to determine due to the low
density of these gas giants, the Rozhen transit has a good time
resolution and thus, the temporal quantities should be considered
with a confidence.

(5) Our values of the relative stellar radius and relative planet
radius are $r_s$=0.1304$\pm$0.0007 and $r_p$=0.01304$\pm$0.00006.
Thus, the obtained values of $a/R_s$ and $R_p/R_s$ are in the
framework of the errors of the previous solutions (Table 2).

\begin{table}[htb]
  \begin{center}
  \caption{Three parameter sets of HAT-P-24b}
  \begin{scriptsize}
  \begin{tabular}{cccc}
\hline\hline
Parameter         & Kipping et al. (2010)       &  Wang et al. (2013)        & this paper       \\
\hline
Period, days       & 3.3552464$\pm$0.0000071     & 3.3552479$\pm$0.0000062    & 3.3552464, fixed       \\
T$_0$, BJD         & 2455216.97669 $\pm$0.00024  & 2455629.67053$\pm$0.00034  & 2455216.97669, fixed        \\
%$T_{14}$          & 0.1539$\pm$0.0008           & 0.1535$\pm$0.0022          &                           \\
%$T_{12}$=$T_{34}$ & 0.0141$\pm$0.0006           &                            &                           \\
$a/R_s$           & 7.70$\pm$0.35               & 7.34$\pm$0.35              & 7.67$\pm$0.21   \\
$R_p/R_s$         & 0.0970$\pm$0.0012           & 0.10625$\pm$0.0126         & 0.1000$\pm$0.0009  \\
\emph{i}, degrees  & 90.0$\pm$1.9                & 88.217$\pm$0.705           & 90.0$\pm$0.1    \\
%$T_{eq}$/$T_{p}$  & 1637 $\pm$ 42               & 1637 $\pm$ 36              & $\leq$1300      \\
%$LD_{lin}$        & 0.1858                      &                            & 0.58                  \\
%$LD_{quadr}$      & 0.3625                      &                            &                   \\
\hline\hline
  \end{tabular}
  \label{table1}
  \end{scriptsize}
  \end{center}
\end{table}

(6) Using the known values of $a$ and masses of the system
components (Kipping et al. 2010) we calculated $R_s$=1.322
R$\odot$, $R_p$ = 0.1322 R$\odot$ = 1.316 R$_J$ and
correspondingly $\rho_p$ = 0.37 g cm$^{-3}$. Our value of planet
radius of HAT-P-24b is by 5 $\%$ bigger than that of Kipping et al
(2010) but almost by 4.5 $\%$ smaller than that of Wang et al
(2013).

(7) We found solutions for different limb-darkening laws which
have almost the same good quality (Table 3) but the best model
(minimum $\chi^2$) corresponded to both quadratic and logarithmic
limb-darkening laws with approximately equal limb-darkening
coefficients (Table 3). Moreover, the freely varying of the
limb-darkening coefficients led us to values appropriate for the
stellar temperature of HAT-P-24 while those of Kipping et al.
(2010) (linear 0.1858 and quadratic 0.3625) differ considerably
from the expected ones.

\begin{table}[htb]
  \begin{center}
  \caption{Limb-darkening coefficients $u_1$ and $u_2$ of the best fits of the Rozhen
  transit corresponding to different limb-darkening laws}
  \begin{scriptsize}
  \begin{tabular}{cccc}
\hline\hline
Limb-darkening law& $u_1$           &  $u_2$        & $\chi^2$       \\
\hline
linear            & 0.490$\pm$0.005 &               & 0.0004484      \\
quadratic         & 0.47$\pm$0.01   & 0.05$\pm$0.01 & 0.0004477       \\
squared-root      & 0.48$\pm$0.01   & 0.02$\pm$0.01 & 0.0004482       \\
logarithmic       & 0.48$\pm$0.01   & 0.02$\pm$0.01 & 0.0004477       \\
\hline\hline
  \end{tabular}
  \label{table1}
  \end{scriptsize}
  \end{center}
\end{table}

\section*{Conclusions}

We presented data of newly observed transit of HAT-P-24b with a 2
m telescope. Its solution gives system parameters which values are
between those of the previous two solutions. The shape and depth
of the new transit favor the maximum orbital inclination of
90$^\circ$.

We obtained transit solutions with the same best quality for four
different limb-darkening laws. The freely-varied limb-darkening
coefficients for these laws turned out almost the same and their values
corresponded to the stellar temperature of HAT-P-24.

We found no evidences of TTV and TDV signals. Hence, until now
HAT-P-24b has no detectable planetary companions on nearby orbits.
This conclusion is in line with the result that the most hot
Jupiters are alone or accompanied by other planetary bodies on
wide orbits.

\section*{Acknowledgments}

This research was supported partly by funds of the project
RD-08-244 of Shumen University. It used the SIMBAD database,
operated at CDS, Strasbourg, France, USNO-B1.0 catalogue
(http://www.nofs.navy.mil/data/fchpix/), and NASA's Astrophysics
Data System Abstract Service.

\end{document}